\documentstyle[epsfig]{mn}

\title[Time-resolved ultraviolet spectroscopy of the compact interacting
binary QU Car] {Time-resolved ultraviolet spectroscopy of the compact
interacting binary QU Car\thanks{Based on observations with the NASA/ESA Hubble Space Telescope obtained at the
Space Telescope Science Institute, which is operated by the Association of
Universities
for Research in Astronomy, Incorporated, under NASA contract NAS5-26555}}
 
\author[L. E. Hartley, J. E. Drew, K. S. Long]{L. E. Hartley$^1$, 
	J. E. Drew$^1$, K. S. Long$^2$\\
$^1$Astrophysics Group, Department of Physics, Imperial College,
Prince Consort Road, London SW7 2BW\\ $^2$Space Telescope Science
Institute, 3700 San Martin Drive, Baltimore, MD 21218\\}

\newcommand{\IV}{\mbox{\thinspace{\sc iv}}}
\newcommand{\II}{\mbox{\thinspace{\sc ii}}}
\newcommand{\III}{\mbox{\thinspace{\sc iii}}}

\newcommand{\V}{\mbox{\thinspace{\sc v}}}

\newcommand{\IVl}{\mbox{\thinspace{\sc iv}}$\,\lambda\,$}
\newcommand{\IIl}{\mbox{\thinspace{\sc ii}}$\,\lambda\,$}
\newcommand{\IIIl}{\mbox{\thinspace{\sc iii}}$\,\lambda\,$}

\newcommand{\Vl}{\mbox{\thinspace{\sc v}}$\,\lambda\,$}

\newcommand{\iue}{{\it IUE}}
\newcommand{\kms}{\mbox{\thinspace km\thinspace s$^{-1}\;$}}
\newcommand{\funits}{\mbox{\thinspace erg\thinspace s$^{-1}\,$cm$^{-2}\,$\AA$^{-1}\,$}}
\newcommand{\hst}{{\it HST}}

\begin{document} 
\maketitle

\begin{abstract}
We present \hst\//STIS (1160--1700\thinspace\AA) echelle spectra of
the cataclysmic variable (CV) star, QU Car, at three epochs. In
catalogues this binary is classified as a nova-like variable. QU Car
was observed three times in time-tag mode for 2300\thinspace sec,
2600\thinspace sec and 2600\thinspace sec, allowing us to study the
spectral time evolution on timescales down to $\sim$10\thinspace
sec. We find evidence of a high-state non-magnetic CV at low
inclination, with unusually high ionisation.

We observed narrow absorption lines ($\sim$ few hundred \kms wide) in
N\Vl1240, O\Vl1371 and Si\IVl1398, as well as broader (HWZI
$\sim1000\kms$) emission in C\IIIl1176, C\IVl1549 and He\IIl1640, all
with a superposed absorption component. High ionisation is indicated
by the He\II\ emission, which is unusually strong in comparison with
C\IV, and the relative strength of the O\V\ absorption line. The
dereddened UV continuum spectral index of, on average, $-2.3$ suggests
that disc accretion dominates the spectral energy distribution. In two
observations velocity shifting is noted in the absorption lines on a
timescale long enough not to repeat within the $\sim2600$-sec
exposures. The absorption superposed on the C\IV\ emission line moves
coherently with the N\V\ and Si\IV\ absorption, suggesting the same
origin for all absorption lines -- most likely to be in the accretion
disc atmosphere.

Weak blueshifted absorption in N\V\ and C\IV\ provides evidence of an
outflow component and we estimate a maximum outflow velocity of
$\sim2000\kms$. This may be linked to a wind launched from further out
in the disc than is typically seen in those high-state non-magnetic CV
whose wind speeds are observed to reach to $\ga4000\kms$. Unusually,
three ionisation stages of carbon -- C\II, C\III\ and C\IV\ -- are
present in emission, with line width increasing with higher
ionisation. The presence of C\II\ in emission and the positive
line-width/ionisation correlation is most easily reconciled with an
origin in a disc chromosphere, beyond the influence of the
EUV-emitting inner disc.
\end{abstract}

\nokeywords
\begin{keywords}
binaries: close -- stars: mass-loss -- novae, cataclysmic variables --
ultraviolet: stars -- line: profiles -- stars: individual: QU Carinae
\end{keywords}

\section{Introduction}

Cataclysmic Variable (CV) stars are interacting binary systems in which a
late-type secondary star overflows its Roche lobe and accretes matter
onto a compact primary. Amongst these, the nova-like variables are
defined as those that persist in a high mass-accretion rate. A result of such 
a heterogeneous label is that, amongst the nova-like variables, there are 
many objects that, displaying similarities to members of other CV classes, 
sit uneasily with their classification. In this paper we present high 
time-resolution UV spectroscopy of one such example -- QU Car.

At $m_{\rm v}=11.4$, QU Car is one of the brightest known CV, so it
is rather surprising that until now it has been mostly ignored in
observational studies. From among the small number of hitherto
published studies, Gilliland \& Phillips
\shortcite{82gilliland} examined optical spectra
(4300--4800\thinspace\AA) at time-resolutions down to 3 minutes. They
derived a long period (10\fh9) from radial velocity variations in the
optical emission lines. They also made note of the relative weakness
of the Balmer line emission, which they deemed due to a high
temperature and high mass accretion rate in the disc and suggested a
classification of old nova or nova-like variable -- a view in keeping
with Schild's (1969) comment that the object displayed some
characteristics indicative of an old nova. Since this time, QU Car has
been classified in catalogues (Ritter \& Kolb 1998, Downes et
al. 2001) as a nova-like variable.

{\it International Ultraviolet Explorer} (\iue\/) low-dispersion spectra
reveal broad blueshifted UV absorption lines in N\V\ and
C\IV, with an emission component in C\IV\ and He\II\ (Knigge, Woods \&
Drew 1994), indicative of an optically thick disc accompanied by mass
outflow. Knigge et al. (1994) noted variation in the C\IV\ line caused
by a periodic increase in the strength of the emission component, but
were unable to decide whether this variablity was linked to the
orbital phase. They also commented on the strength of the
O\Vl1371 absorption line and the He\IIl1640 emission -- both unusually
strong for high-state non-magnetic CV (HnMCV).

The data that are analysed in this paper were obtained as part of an
observing campaign to detect spectral signatures of disc-wind
variability of short timescales. These signatures are predicted to
show up as highly time-variable fine structure components in wind-formed 
UV lines (Proga, Stone \& Drew 1998). To achieve the necessary time and 
wavelenght resolution, the Space Telescope Imaging Spectrograph (STIS) on the
{\it Hubble Space Telescope} (\hst\/) was employed with an echelle
grating in photon-counting (TIME-TAG) mode. This allows a
time-resolution of down to $\sim10\,$s and velocity resolution of down
to $\sim10$\kms, with the trade-off that higher time-resolution will
result in a reduced useable spectral resolution and vice-versa.

On examining the results of the \hst\/ observations we were struck by
the ways in which QU Car distinguished itself from the other targets in
the programme (the nova-like variables IX Vel and V3885 Sgr, presented in 
Hartley et al. 2002). In this paper we seek to present the current \hst\/
observations and then re-examine archive data, with a view to
establishing a physical origin for the unusual features noted. In
section \ref{s:obs} we describe the observations and data extraction
processes. The results are then presented in two ways: in section
\ref{s:mean} we describe the time-averaged spectrum obtained at each
epoch of observation, then in section \ref{s:timevar} we re-present
the data as spectral time-series at 30-sec time-resolution and as
continuum light curves. Finally, in section \ref{s:discuss} we 
complare QU~Car with other non-magnetic HnMCV and consider the origin
of QU~Car's UV line spectrum.

\section{Observations and Data Extraction}
\label{s:obs}

Observations were performed by the STIS instrument on the \hst\, with
the far ultraviolet (FUV) MAMA detector (1140--1735\thinspace\AA). The
E140M echelle grating with a central wavelength of 1425\thinspace\AA\
was used. This configuration provides a resolution of
$\sim13\kms$ (R $\sim$ 23000). Each observation was performed in a single 
telescope orbit for the maximum available time, with the detector set to
time-tag mode and using the 0.2\,\arcsec$\times0.2\,$\arcsec\
aperture. The dates and exposure lengths of our observations are given
in table \ref{tab:obs_dates}.

\begin{table}
\label{tab:obs_dates}
\begin{tabular}{@{}lccc}
\hline\hline
Observation date	&Label	&U.T. start time  	&Exposure\\
(2000)		&	&		&(s)	\\
\hline
12th Feb		&Q1	&20:26:07	&2300	\\
24th July		&Q2	&07:13:34	&2600	\\
16th Sep		&Q3	&02:56:32	&2600	\\
\hline
\end{tabular}
\caption{The observation dates and times for each observing run}
\end{table}

All data calibration was performed with {\sc iraf} software, using the
{\sc stsdas} package produced by the Space Telescope Science Institute
(STScI). Time-tag is a photon counting mode, which provides an events
stream with 125-$\mu$s time-resolution. This can be integrated over any
selected time resolution (time bin) to produce a set of raw images of
the echelle output, from which a 1-D spectrum is extracted and
calibrated. The calibration, including doppler correction for the
telescope's motion, was performed with STScI's {\sc calstis} package,
using the suggested best reference files.

\section{The mean UV spectrum}
\label{s:mean}

\begin{figure*}
\vspace*{10cm}
\includegraphics{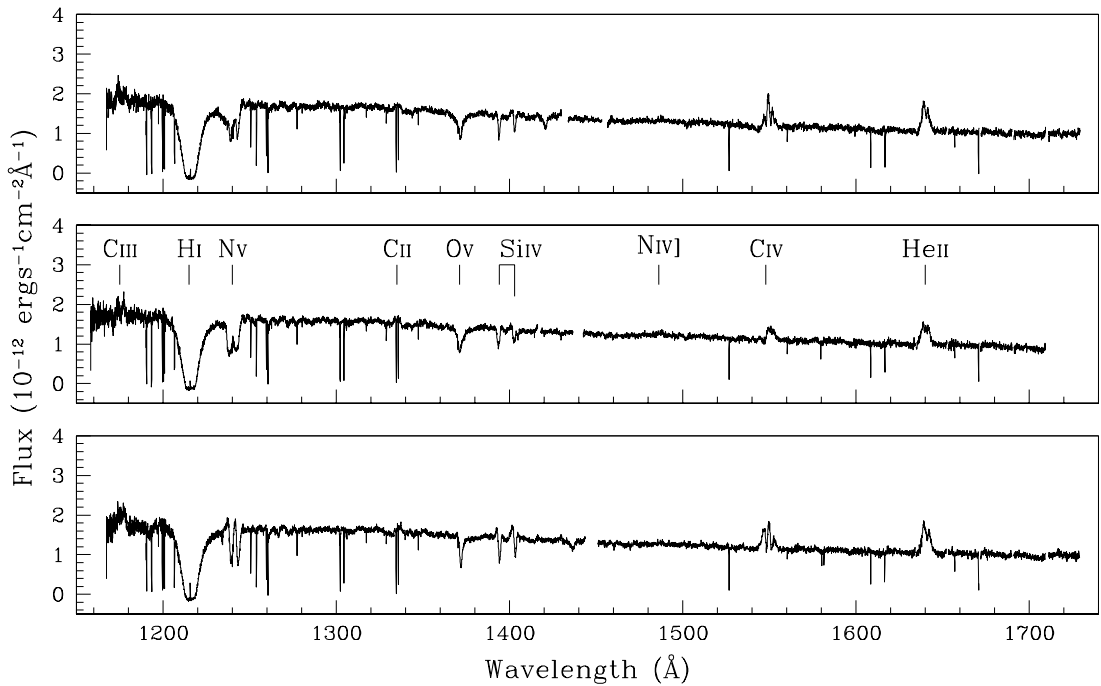}
\caption{The summed spectrum for (from top to bottom) 12th January, 21st July and 16th September 2000. Resonance transitions and other strong lines are indicated.}
\label{fig:qucar}
\end{figure*}

figure~\ref{fig:qucar} shows the summed spectrum for each observation
(hereafter called datasets Q1, Q2 and Q3). QU Car's UV lines appear to
be characterised by strong absorption features in N\Vl1240 and
O\Vl1371, less prominent Si\IVl1398 absorption, and emission in
C\IIIl1176, C\IVl1549 and He\IIl1640. The narrowness of QU Car's line
features is striking. Typical terminal velocities of outflows in
high-state CV that commonly determine UV line widths can be up to
5000\kms.  Indeed it is usual to find broad blueshifted absorption in
the stronger resonance line profiles in non-eclipsing systems. Here,
however, the line profiles rarely extend to more than
$\sim\pm1000$\kms, and show rather little P Cygni behaviour.

The STIS datasets offer an insight into the finer structure of lines
that have previously only been identifiable in \iue\/ low-resolution
data. However, before describing the character of the observed profiles,
we first consider whether it is possible to correct the spectra for line of 
sight (LOS) motion of the white dwarf (WD). Gilliland \& Phillips
\shortcite{82gilliland} calculated a spectroscopic zero-phase
ephemeris of $T_0=2443960.683\pm0.003$JD and a period of
$10\fh896\pm0.336$ from velocity shifts in the He\IIl4686 emission
line. However, the uncertainty in the phase measurement is such that
we are unable to determine the orbital phase of the data
presented here. So no radial velocity correction can yet be applied to
the spectral lines. Nevertheless, the orbital $K$-velocity ($K_{\rm
WD}=115\pm13$\kms) and radial velocity ($\gamma=-84\pm20$\kms) derived
by Gilliland \& Phillips suggest the data are radial velocity shifted
from the WD rest frame by no more than $\sim200$\kms bluewards and
$\sim30$\kms redwards.

\begin{figure*}
\vbox to220mm{
\includegraphics{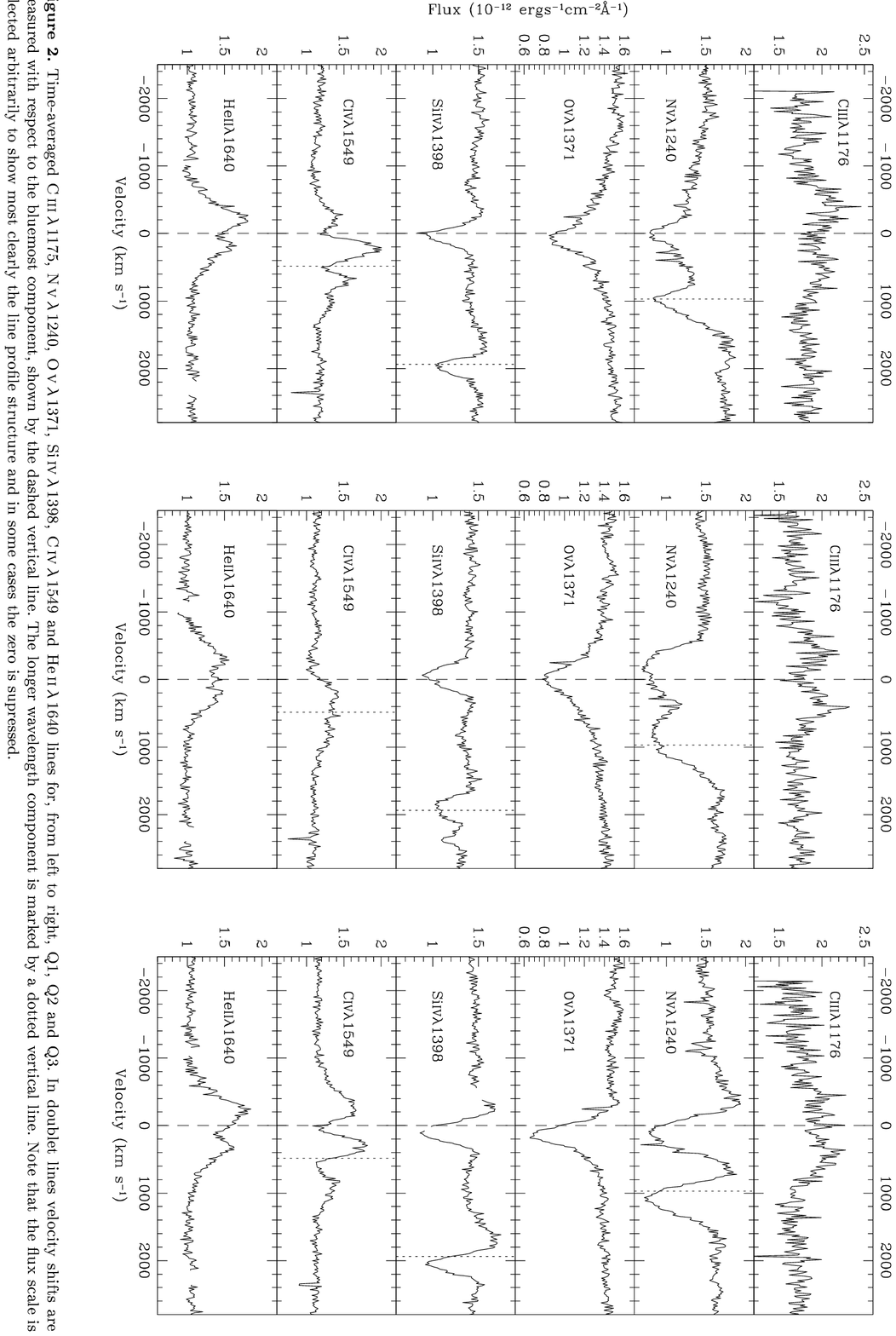}
}
\label{landfig1}
\end{figure*}
\setcounter{figure}{2}

Figure 2 presents time-averaged close-ups of the strongest lines
apparent in QU Car's UV spectrum, at each epoch of observation, with the 
wavelength given in the heliocentric frame.  

N\V$\,\lambda\,\lambda\,$1238.8,\thinspace1242.8 doublet is the
transition that shows the most secular variation. In Q1 the gross
structure of the N\V\ profile bears the mark of an outflow component
with a broad asymmetric absoption trough that stretches from
approximately $-2000$ to $+1500$\kms, relative to the bluemost
component of the doublet ($+500$\kms relative to the red
component).
Superposed on this broad structure are narrower absorption troughs,
$\sim400$\kms FWHM, centred on the doublet rest wavelengths. In Q2 the
N\Vl1240 line is in absorption from $-600$ to $\sim+1500$\kms, with
each component of the doublet clearly defined, but overlapping
slightly and more rounded than in the other sets. In Q3 there is
clearly blueshifted emission in N\V\ (to no more than $-1000$\kms) and
the absorption minumum is redward of heliocentric rest at about
$+100$\kms.  Without knowing the exact LOS velocity correction we
cannot be certain that the absorption minimum is redshifted in the
primary frame - such a conclusion rests heavily on the adopted $gamma$
velocity. If the maximum likely white-dwarf (primary) redshift really
is no greater than $\sim30$\kms~(see above), the absorption minimum
remains at net redshift.

During the first epoch of observation, Q1, the O\V\ line is broad with
absorption reaching to at least $\pm1500$\kms. The profile shows a
skew that suggests stronger absorption in the blue wing, indicating
some contribution from an outflow.  This transition is remarkably
strong compared to other nova-like variables in which it is seen as a
weak, rotationally-broadened, disc-formed line, if seen at all
(c.f. the HST/STIS data on IX~Vel and V3885~Sgr presented in Hartley
et al 2002)). In all three datasets the Si\IVl1398 absorption profile,
seen in both doublet components, is about $250\kms$ wide (FWHM).
There is no blueshifted absorption at any time analogous to that seen
in N\V\ and weakly in O\V .

The changes seen in Si\IVl1398 and O\Vl1371 are similar, but less
marked than those seen in the N\V\ line -- this lends a general pattern to 
QU Car's core line absorption: In Q1 we see fairly symmetrical, almost 
triangular absorption; in Q2 the lines are somewhat more rounded; in 
Q3 there is a definite redwards skew to the absorption lines, with blueshifted
emission components obvious in N\V\ and Si\IV.

The C\IIIl1176, C\IVl1549 and He\IIl1640 lines are in emission to
about $\pm1000$\kms. The C\IV\ and He\II\ emission lines are
respectively triple- and double-peaked with peak separations of
$\sim400$\kms.  Significantly, the absorption minima in C\IV\ are
separated by the intrinsic doublet splitting of 484\kms (to within
measurement errors).  In all observations the C\III\ line's emission
peaks are split by 800\kms -- the individual components of this
six-component transition span 370\kms.  The most plausible
interpretation of the profile morphology is that, in each case, a
broad (FWHM $\sim1000$\kms) emission feature is cut into by narrower
absorption, broadened kinematically to only FWHM $\sim400$\kms.  The
shapes of the absorption components resemble the shapes of the
absorption-dominated N\V\ , O\V\ and Si\IV\ lines discussed above. The
physical origin of the line spectrum will be considered later in
section
\ref{s:discuss}.

We note little secular variation in the emission lines, although, Q2
does set itself apart from the other two observations. Firstly, the
emission lines are weaker than in the other datasets, whilst there is 
not much corresponding weakening of the superposed absorption components. 
Also, in Q2, the C\IV\ line has lost its triple-peaked shape, becoming 
instead what appears as a weak P~Cygni form with redshifted emission 
from $-200$ to $+1400$\kms\ and a hint of blueshifted absorption 
also. 

Before making a fit to the spectral energy distribution, the spectrum
was dereddened using $E(B-V)=0.1$ given in Verbunt (1987).
A power law continuum fit was made to the mean spectrum, after masking
out areas obviously contaminated by spectral lines. Table \ref{tab:cont} 
lists the spectral index measured from each mean spectrum and the average 
flux level in the 1260--1270\thinspace\AA\ range. We note very little 
variation in the continuum flux from observation to observation. QU Car's 
spectral index over the $\sim$500~\AA\ available to us is $-2.4$ on average, 
fitting in well with other nova-like variables, for example IX Vel (-2.4) 
and V3885 Sgr (-2.3) (Hartley et al. 2002).  Corrected for the same assumed
reddening, merged short and long wavelength IUE spectra of QU Car spanning 
nearly 2000~\AA\ fit to a power law index of -2.25 $\pm$ 0.1 (see Drew
et al. 2002).  This suggests the UV spectral energy distribution is 
reasonably stable over time.

\begin{table}
\caption{Mean continuum flux level in the range 1270--1320
\thinspace\AA, with  mean continuum flux and index of power law continuum fit, when the spectrum
is dereddened with $E(B-V)=0.1$. Flux units
are $\times10^{-12}$\funits. Error is one standard deviation from the
mean.}
\label{tab:cont}
\begin{tabular}{@{}lccl@{}}
\hline\hline
&&\multicolumn{2}{c}{Dereddened}\\
Label &Flux &Flux&Index \\
\hline
Q1	&$1.66\pm0.13$	&$3.88\pm0.12$&$-2.32\pm0.05$\\
Q2	&$1.56\pm0.13$	&$3.65\pm0.13$&$-2.45\pm0.10$\\
Q3	&$1.61\pm0.13$	&$3.75\pm0.10$&$-2.24\pm0.08$\\
\hline
\end{tabular}
\end{table}


\section{Variation on timescales shorter than $\bmath{\sim 2500}\,$sec}
\label{s:timevar}

An analysis of time-variability of the strong lines during each
observation is presented in this section. The N\Vl1240, O\Vl1371,
Si\IVl1398 and C\IVl1549 lines are displayed as trailed
mean-subtracted spectra, as trailed normalized spectra and as 1D
averages of the first and the second half of the 2600-sec
observation. In the trailed spectra we include only observations Q2
and Q3, as Q1 displays no sign of time variability during our
observation (see below).

To reduce the prominence of noise in the trailed spectra, the 1D
spectra were first smoothed by convolution with a gaussian function of
sigma, $\sigma=1.1\,$\AA\ (10 pixels). The spectra were then normalized
by fitting a 4th order chebyshev function to the time-averaged data;
the fit was checked to assure that it did not produce strong curvature
in the vicinity of the spectral lines. All time-resolved spectra were
divided by this continuum fit and then renormalized to ensure a
continuum level of 1. The mean-subtracted spectra were prepared for
plotting by subtracting the normalized time-averaged spectrum from
each time-bin's normalized spectrum.

\begin{figure*}
\vspace*{9cm}
\includegraphics{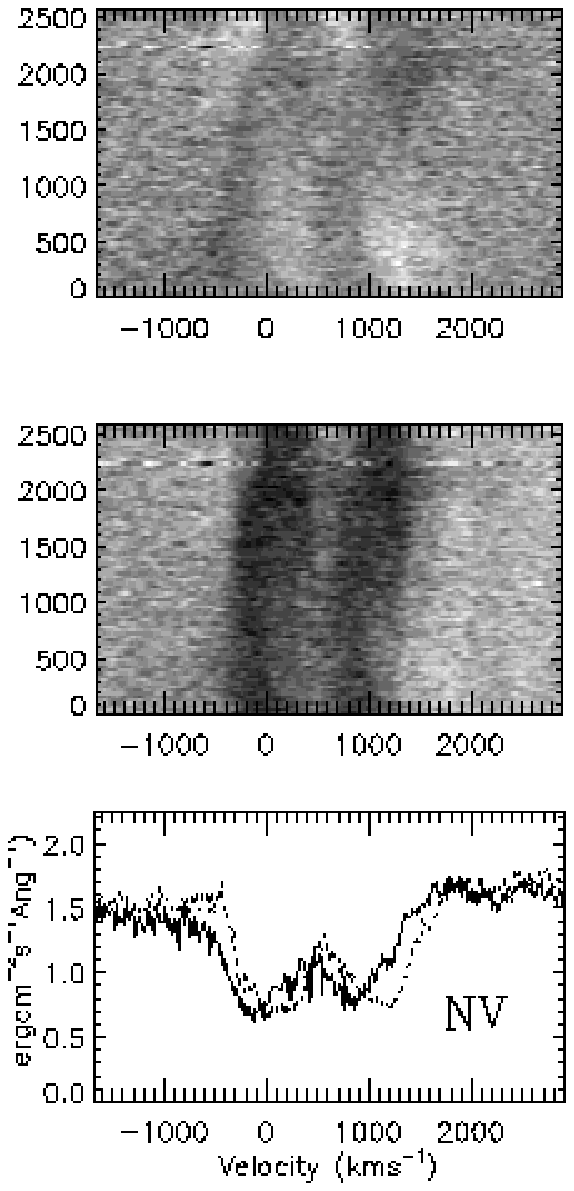}
\includegraphics{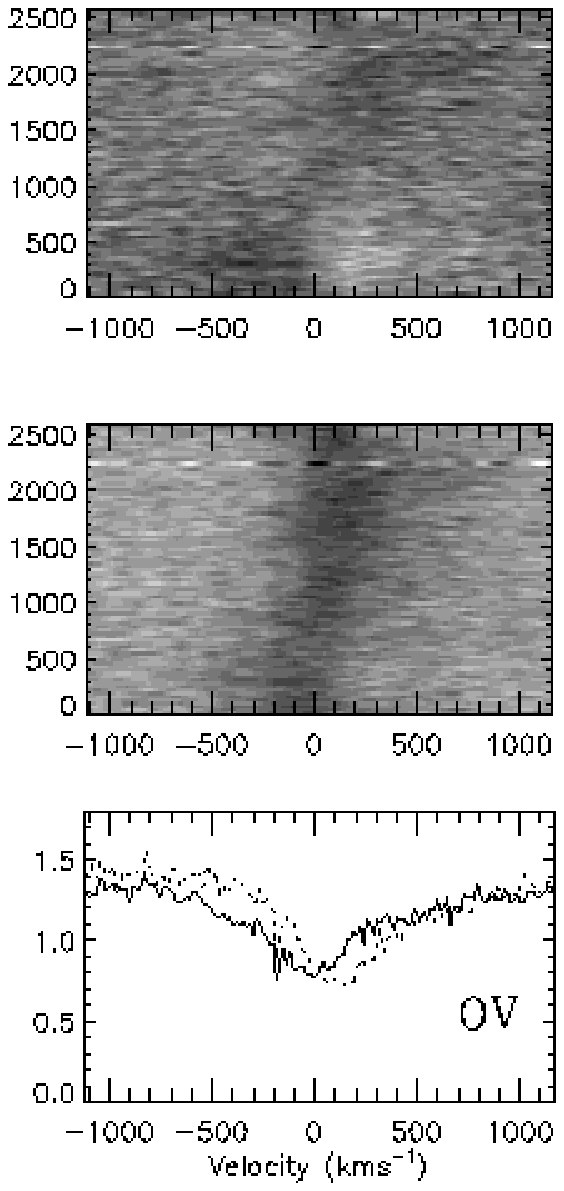}
\includegraphics{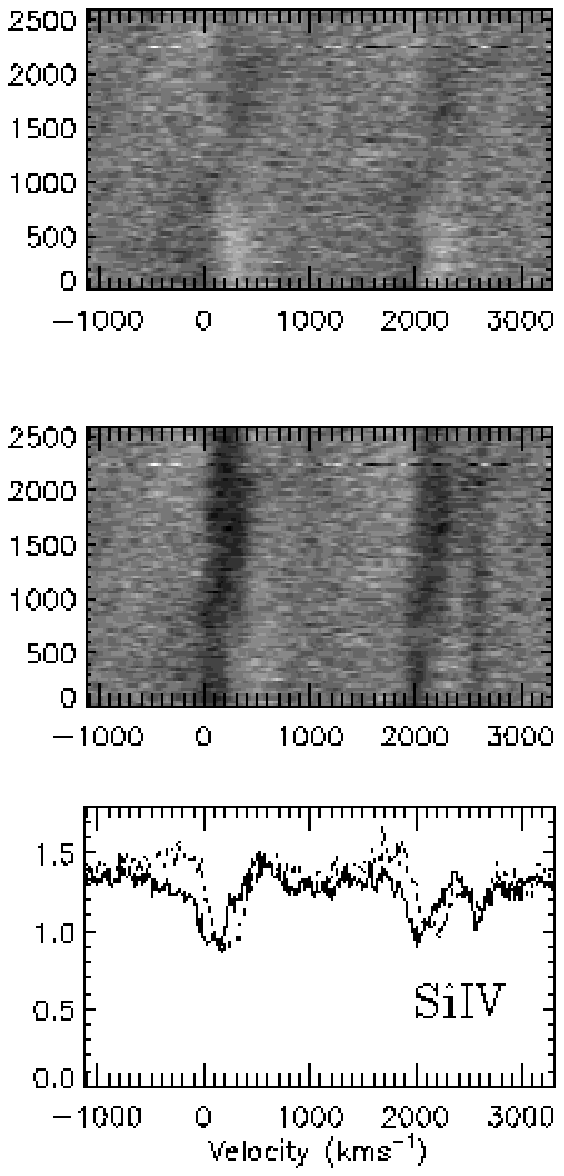}
\includegraphics{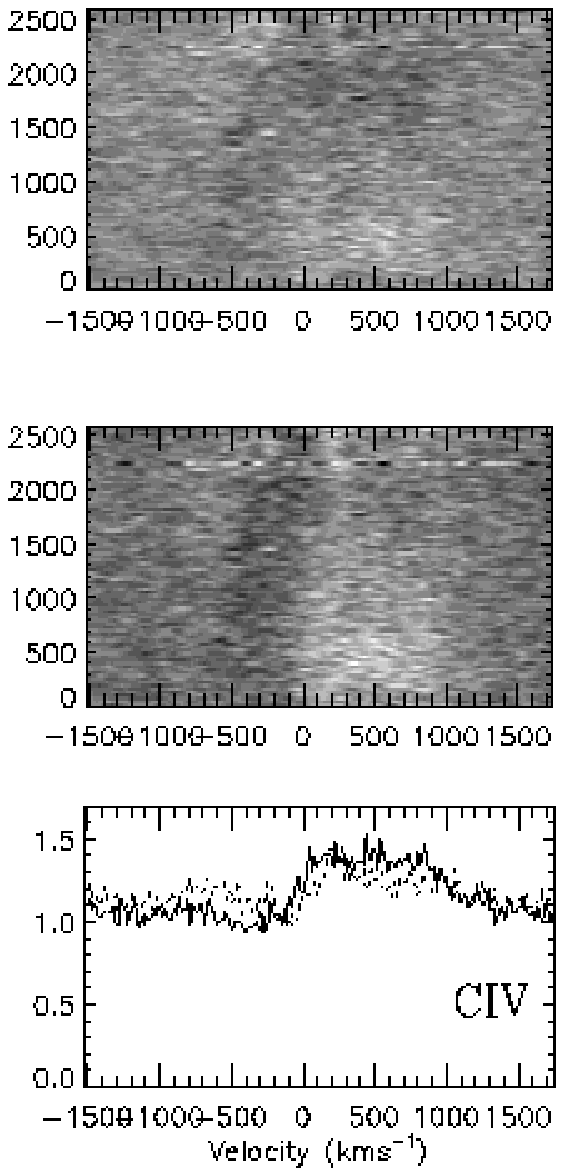}
\caption{Mean-subtracted difference spectrum of (left to right, top to
bottom) the N\Vl1240, O\Vl1371, Si\IVl1398 and C\IVl1549\ lines for Q2. The flux range is as
follows (in mean-subtracted normalised flux units): N\V, -0.28 to
0.30; O\V, -0.26 to 0.33; Si\IV, -0.25 to 0.34; C\IV, -0.44 to
0.49. Plotted beneath the stacked spectrum is the mean line profile for
the first half of the observation (solid line) and the second half
(dotted line).}  \label{fig:gr2}
\end{figure*}

\begin{figure*}
\vspace*{9cm}
\includegraphics{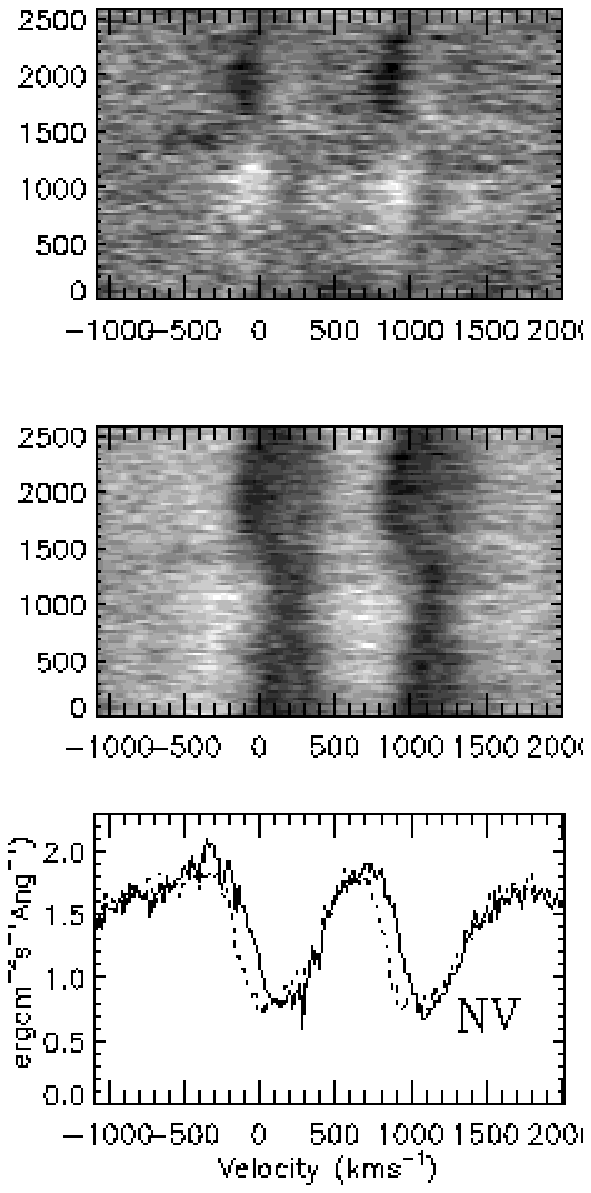}
\includegraphics{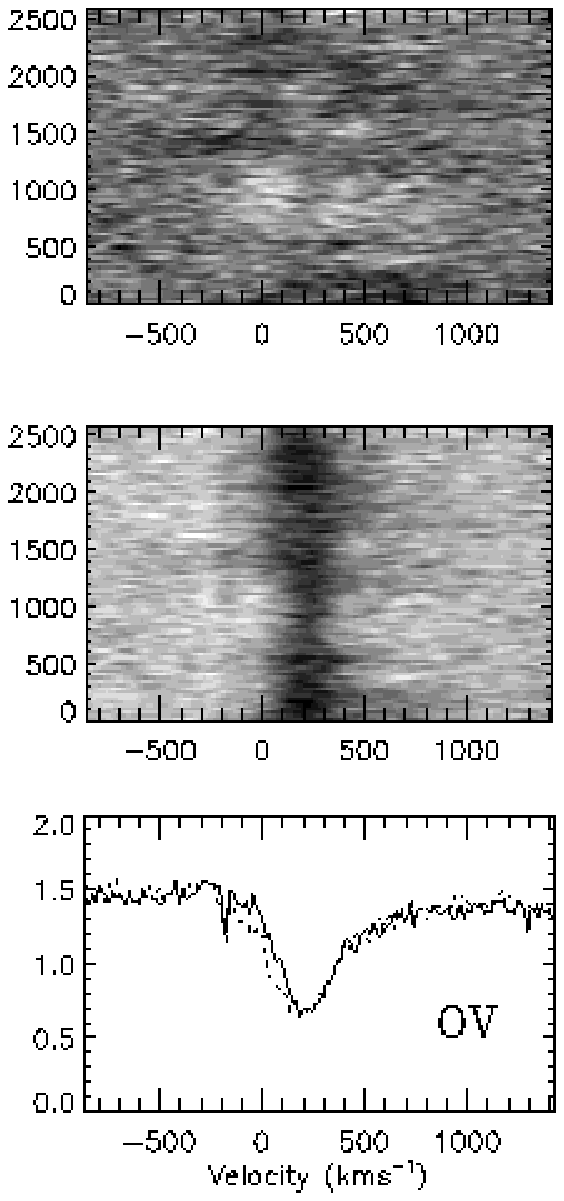}
\includegraphics{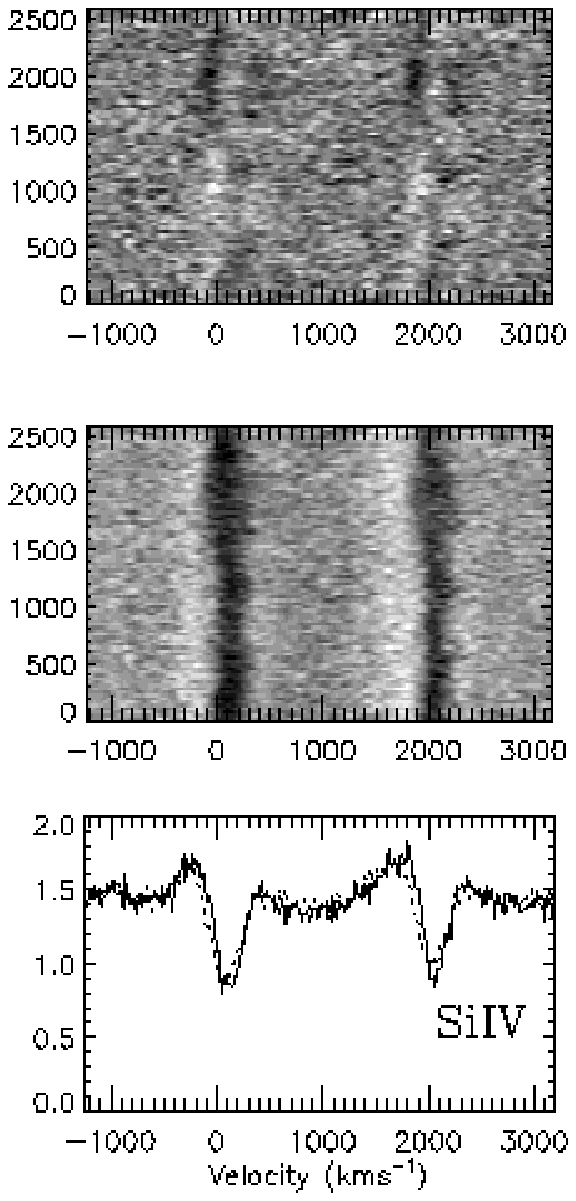}
\includegraphics{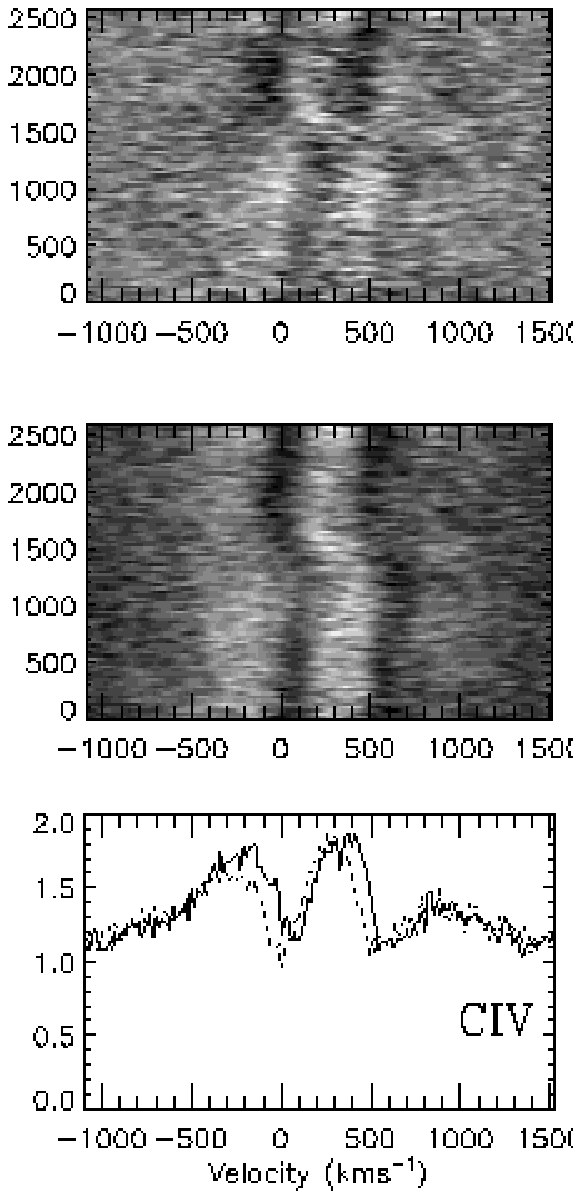}
\caption{Greyscale difference spectrum of (left to right, top to
bottom) the N\Vl1240, O\Vl1371, Si\IVl1398 and C\IVl1549 lines for Q3. The flux range is as
follows (in mean-subtracted normalised flux units): N\V, -0.32 to
0.39; O\V, -0.31 to 0.27; Si\IV, -0.28 to 0.25; C\IV, -0.40 to
0.46. Plotted beneath the stacked spectrum is the mean line profile for
the first half of the observation (solid line) and the second half
(dotted line).}  \label{fig:gr3}
\end{figure*}

\begin{figure}
\vspace{12cm}
\includegraphics{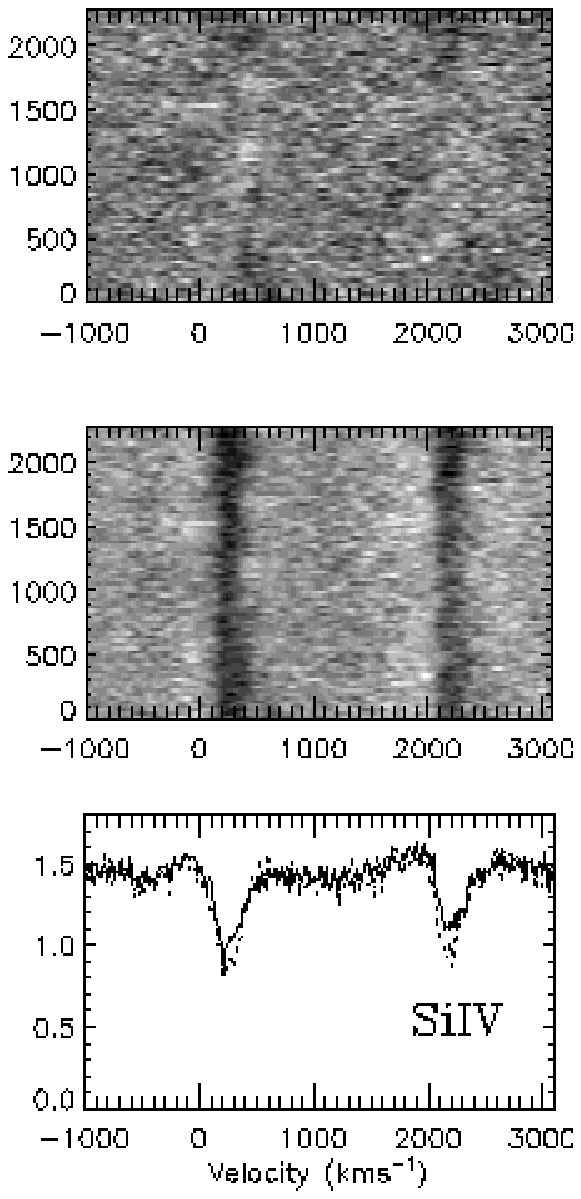}
\caption{Greyscale difference spectrum of the Si\IVl1398 line for Q1.
The flux range is -0.22 to 0.24
(in mean-subtracted normalised flux units). Plotted beneath the
stacked spectrum is the mean line profile for the first half of the
observation (solid line) and the second half (dotted line).}
\label{fig:gr1}
\end{figure}

The Q2 trailed spectra are presented in figure
\ref{fig:gr2}. In N\V\ a dark broad  ($\sim500$\kms
across) feature shifts steadily to the red by about 500\kms.
Examining the 1D spectra for the first and the second half of the
observation separately, it becomes clear that this is due to a
redwards shift in the entire absorption profile. This is too large to
be accounted for by a radial velocity shift due to WD orbital
motion, if the orbital parameters derived by Gilliland and Phillips 
(1982) are accepted.

A similarly broad dark feature is also seen in the Si\IV\ and O\V\
trailed spectra. However, compared to N\V\ the shift is more abrupt at
around the midpoint of the observation, fliping from blue to red in
the space of less than 500 sec. Again, this feature represents a shift
in the entire absorption profile from blue to red.  Along with the
redward shift there is also an alteration in the form of the O\V\ and
Si\IV\ profiles, with the bluemost absorption wing becoming weaker and
possibly even slightly emissive in Si\IV. If more subtly, the C\IV\ line 
appears to shift redwards in the same way as the N\V\ line. Also, its 
broad redshifted emission component weakens slightly during the observation. 
No variation is detected in either the C\IIIl1176 or He\IIl1640 lines.

In the Q3 trailed spectra (figure~\ref{fig:gr3}) there appears to be a 
velocity-shifting component to the line profiles: in the N\V, Si\IV\ and 
C\IV\ lines a dark feature ($\sim$200\kms wide) shifts gradually redwards by 
$\sim200-300$\kms and then more sharply bluewards around the midpoint of the 
observation; this component then resumes its redwards shift, creating a 
`sawtooth' pattern. As this variation occurs, the redward edge of these line 
profiles is very nearly stationary (see the lower panels in 
fig.~\ref{fig:gr3}).  There is at most just a hint of the same behaviour in 
the O\V\ line profile -- the most obvious change is a slight weakening
and narrowing of the absorption as a whole for about 300~sec centred on
1000~sec after the start of the observation.  Again, the C\III\ and He\II\ 
lines do not vary noticeably.

Q3 does reward with some sporadic, shorter timescale variablity. In
the N\V\ and O\V\ trailed spectra there is a broadening of the
absorption, appearing first in the red wing at around 1150~sec and
then switching into the blue wing, finishing at about 1600\thinspace
sec.  In the C\IV\ trailed spectrum, just the beginning of this change
affecting the red wing of the line is detectable.  There may also be a
weaker event of the same kind at about 600~sec after the start, that
again shows in the N\V\ and O\V\ profiles.

The variability seen in Q2 and Q3 is quite different. In Q2, there appears to 
be a shift in the gross line profiles accompanied by a small change in the 
steepness of the blue absorption edge. In Q3 the blue absorption 
edge shifts further to the blue, yielding an increase in the width of the
line profile and only in this dataset is there any apparent variation on
the $\sim100$~sec timescale. In Q3, where line emission is generally
more prominent, the variability persists as an attribute of the
absorption line spectrum.

The lack of variabilty in observation Q1 is interesting as in the mean 
spectrum it bears a stronger resemblance to Q3 than the more variable Q2. 
Figure~\ref{fig:gr1} shows the Q1 trailed data for Si\IVl1398, the most 
variable of the transitions observed at this time. There is just a 
trace of the velocity-shifting component that is seen in the Q3 Si\IV\ line. 
In the Q1 N\V\ line profile there is a weak component of broad
blueshifted absorption, signalling the presence of an outflow (see section
\ref{s:mean}). At the time of the Q3 observation there was no such evidence
of outflow.  It remains to be seen whether this anticorrelation is a 
persistent and hence significant effect.

\begin{figure}
\vspace{8cm}
\includegraphics{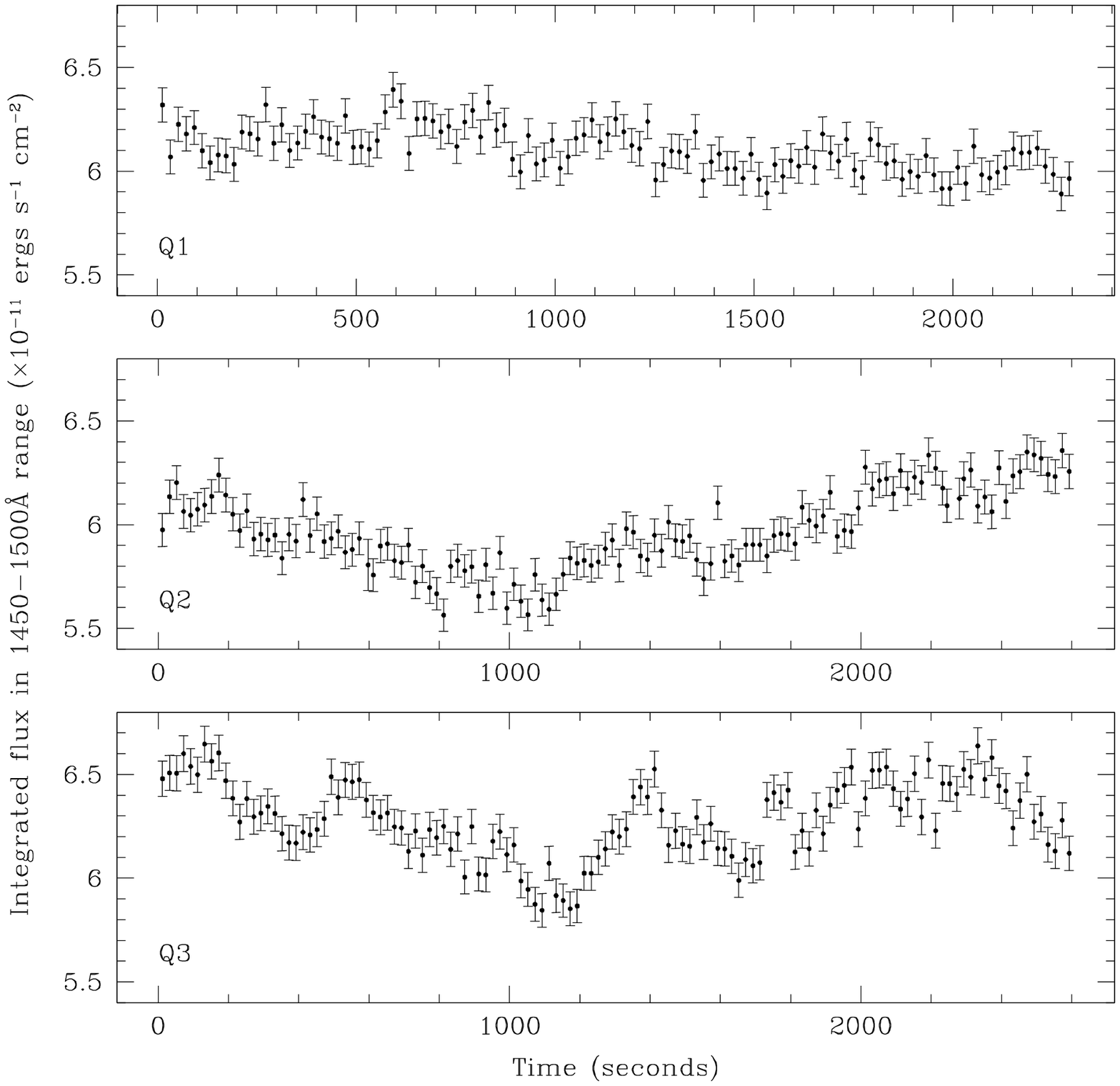}
\caption{Light curve from each observation of QU Car. The flux is integrated over the wavelength
range 1450--1500\AA\ and the time-binning is 20\thinspace s.}
\label{fig:lc}
\end{figure}

In order to examine the data for continuum variability we plotted light-curves
(figure~\ref{fig:lc}) of flux summed over the 1450--1500\thinspace\AA\
wavelength range, with time-binning of 20\thinspace s for each dataset. There 
is a similar contrast in the degree of continuum variability as that seen in 
the line profiles: Q1 exhibits almost no variability; Q2 shows some
brightness change, dimming slightly towards the middle of the
observation, and then recovering; finally, the lightcurve for Q3 has the
most chaotic appearance of the three.  The power spectrum was calculated
for each of the three time series -- in none was there evidence of any 
preferred periods of variation.

The detected UV light variation has an amplitude of around 10 per cent. This 
is consistent with optical estimates of photometric variability.  Schild
\shortcite{69schild} reported optical brightness fluctuations of up to
0.2\thinspace mag on a timescale of 90\thinspace sec.  Similarly
Gilliland and Phillips (1982) and Kern \& Bookmyer (1986) presented
photometric data showing variations of 0.1--0.2\thinspace mag on
timescales of a few minutes.

\section{Comparison with previous \textbf{\it IUE\/} observations}
\label{s:mag}

To set the characteristics of our high-quality \hst\/ UV
spectra of QU Car in a broader context, we retrieved the 15 low-dispersion
($\Delta\lambda=1.1\,$\AA) 1150--1975\thinspace\AA\ spectra available
from the \iue\/ archive. These data are provided pre-calibrated in the
NEWSIPS format and was extracted via the {\sc iuesips} software run
under {\sc iraf}.

The continuum flux level of our \hst\/ data is consistent with, but at
the high end of that recorded in previous \iue\/ observations in
which it varies from a minimum of $\sim1.38\pm0.09\times10^{-12}$ to
maximum of $\sim1.61\pm0.14\times10^{-12}$\funits (cf table~\ref{tab:cont}).  
The \iue\/ observations yield, on average, a slightly flatter dereddened
continuum slope ($-2.24\pm0.13$) than do the \hst\/ data ($-2.34\pm0.07$).

As regards the spectral line behaviour, in contrast to the \iue\/
data, the HST/STIS observations have caught QU Car at times when its 
C\IVl1549 line presents with almost no blueshifted absorption. In all
but three of the \iue\/ spectra there is a significant blueshifted
absorption component to the C\IV\ line, extending to between $-2000$ and
$-2500$\kms of line centre. If this is taken to be the terminal
velocity of an outflow, then this is relatively modest, compared to other
CV, such as IX~Vel and V3885~Sgr (Hartley et al. 2002) where the
C\IV\ line shows evidence of outflows up to at least 5000\kms.  In
N\Vl1240, there is always a blueshifted absorption component.  The 
proximity of the damped Ly$\alpha$ absorption may render this line less 
suitable, at the low resolution of the \iue\ low dispersion mode for a 
determination of typical maximum outflow speed.  Nevertheless, measurement
of the collected \iue\ data and the Q1 \hst\ observation (figure 2) 
indicate that the maximum blueshift in the N\V\ line generally matches 
that in C\IV.

\begin{figure}
\vspace{8cm}
\includegraphics{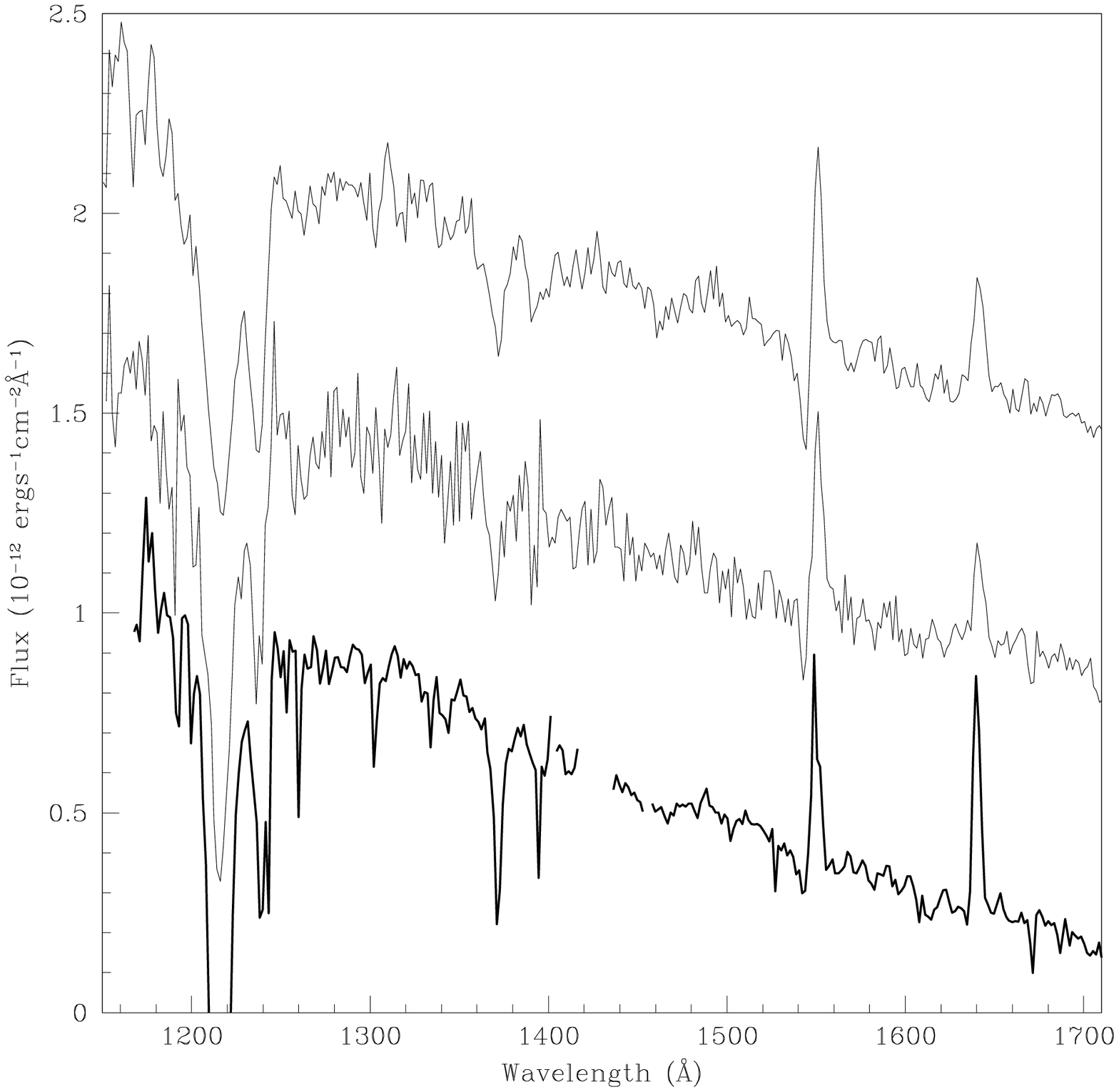}
\caption{Two representative merged \iue\/ spectra of QU Car compared
to the Q1 \hst\/ spectrum downgraded to \iue\/ resolution. The
continuum level was approximately the same in all three
observations. The \iue\/ merged spectra SWP41925-30,
obtained 26/27 June 1991, are plotted on the correct level, but for
clarity, the \iue\/ merged spectra SWP15670-1, obtained 7 Dec 1981,
are offset upwards by $6\times10^{-13}\,$\AA\ and the Q1 \hst\/
spectrum is offset downwards by $8\times10^{-13}\,$\AA.}
\label{fig:iue}
\end{figure}

Figure~\ref{fig:iue} shows dataset Q1 downgraded to \iue\/ resolution
and overplotted on merged short-wavelength \iue\/ data for 1981 December
7 (SWP15670, SWP15671), selected for being most similar to the \hst\/ sets, 
and merged data from 1991 June 26/27 (SWP51925-30), selected for showing
much better developed wind signatures.  We note that the He\II\ emission is 
somewhat stronger in the \hst\/ sets than in all \iue\/ observations.  The 
O\Vl1371 line is present as a clear absorption feature in all UV spectra 
obtained to date.  Putting all the available UV data together, there is no 
evidence that the equivalent width of the blueshifted absorption in the 
C\IV\ and N\V\ lines changes systematically with e.g. UV continuum 
brightness.  Knigge et al. (1994) provide further discussion of variability
in the \iue\ data. 

Apart from the tendency for clearer wind signatures in the N\V\ and
C\IV\ line profiles to appear in the \iue\/ data (particularly those from the
epoch studied by Knigge et al. 1994), there is strong consistency between 
these older observations and the new \hst\ data.  Encouraged by this, we
can now turn to a summary discussion of what the higher-quality \hst\/ 
observations of QU~Car have revealed to us about this highly-ionised 
compact binary.


\section{Discussion}
\label{s:discuss}
In this paper we have presented high time- and high spectral-resolution UV 
spectra of QU Car obtained at three epochs.  These data have pointed toward
a higher than typical degree of ionization, and reveal a mix of fairly 
narrow, emission and absorption line features along with only
modest evidence of mass loss.  

Before considering the physical implications of the new spectroscopy, we 
review the constraints on the inclination of QU~Car.  Gilliland \& Phillips 
proposed $i\la60\degr$ from the lack of orbital modulation in the binary's 
light curve and the relatively narrow optical emission features.  The UV
emission lines are not particularly broad either, generally fitting within
approximately $\pm1000$\kms.  The simple fact that a number of UV spectral
lines are firmly in absorption, and that broadened blueshifted absorption
is detected from time to time, also reinforces the case for low
inclination in that line emission is accepted as the general pattern for 
high-state high inclination systems.  We can venture a little further
than this and note that the observed mix of emission and absorption lines
argues persuasively against re-classifying QU~Car as a magnetic system (see
Warner 1995) -- a step that might otherwise be contemplated on the grounds of 
the strong He\IIl1640 emission typifying this object.

\begin{figure}
\vspace{9cm}
\includegraphics{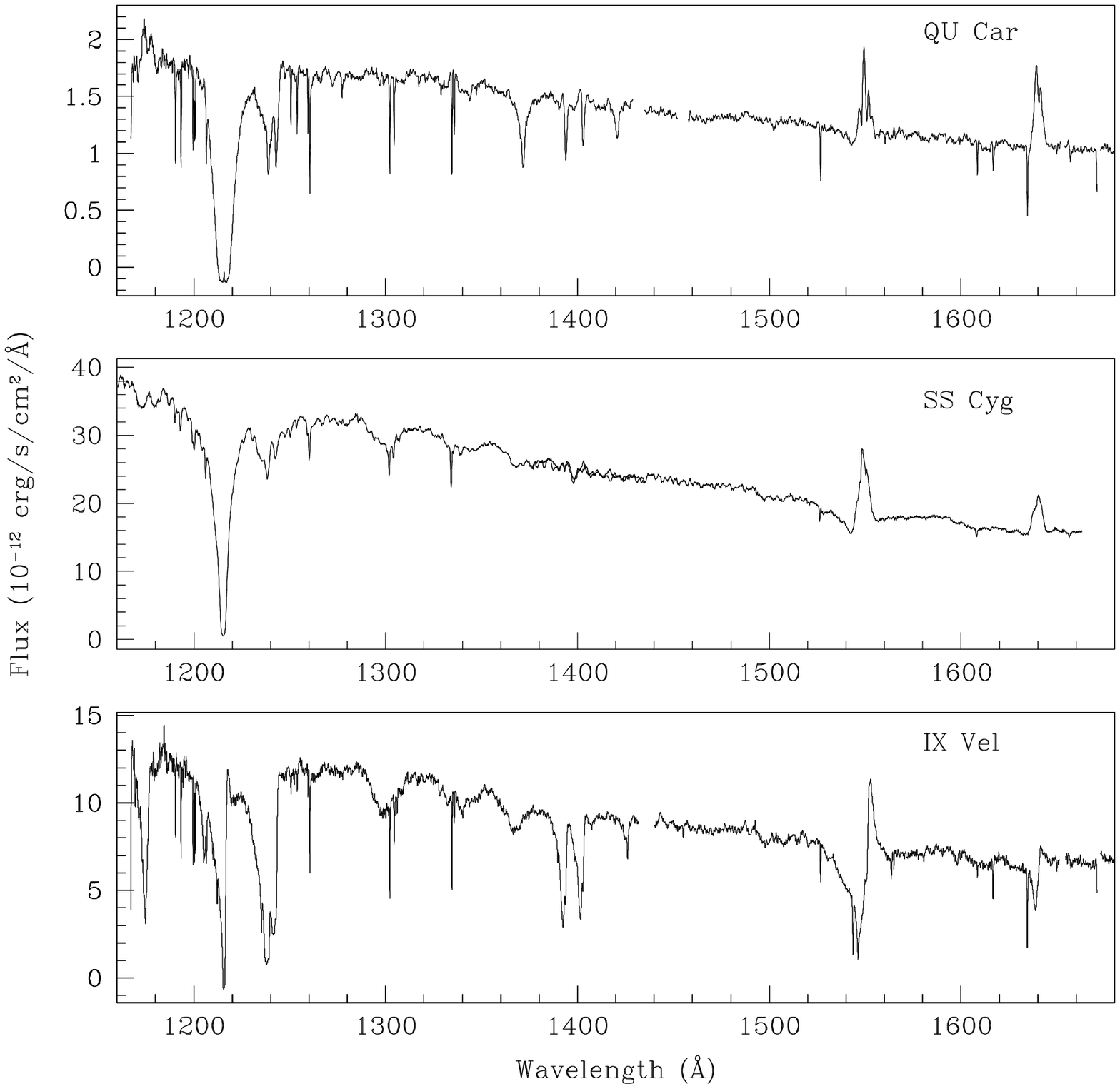}
\caption{STIS spectrum of QU Car (dataset o5bi07010) compared to a GHRS spectrum
 of the dwarf nova SS~Cyg in outburst (datasets Z3DV0104T and
 Z3DV0105T) and a STIS spectrum of the novalike variable IX Vel
 (dataset o5bi01010). All specta were downloaded via the MAST facility
 at the STScI.}
\label{fig:comp}
\end{figure}

We now compare the 1150--1660~\AA\ spectrum of QU Car with those of
more typical high-state non-magnetic CV in order to make the case that
QU~Car is extreme in terms of the apparent degree of ionization.  We
choose as our comparison objects SS~Cyg near peak of outburst and
IX~Vel, for the reason that similarly high quality \hst\ UV spectra
have been obtained for them.  IX~Vel and SS~Cyg are also non-eclipsing
systems, like QU~Car.  The SS~Cyg observations used (Z3DV0104T and
Z3DV0105T) were obtained by one of us (KSL) and, to ensure the most
up-to-date calibration, were retrieved from the MAST facility at
STScI.  The IX~Vel observations were presented in Hartley et al
(2002).  The spectra of these three sources are shown together in
figure~\ref{fig:comp}.  The three objects are placed in a sequence of
increasing degree of ionization, with IX~Vel the least-ionized, QU~Car
the most ionized and SS~Cyg somewhere in between.  This assessment is
based largely on how the He\IIl1640 and O\Vl1371 line profiles compare
with C\IVl1549 -- these signatures of high excitation are indeed
relatively strongest in QU Car.

Anther point of contrast between QU~Car and the other two systems concerns 
the broad absorption dips at $\sim$1300~\AA\ and $\sim$1340~\AA . These
are presumably due to disc atmospheric blanketing.  Wade \& Hubeny (1998) 
show in their disc atmosphere models that these features would be expected to 
fade as the maximum disc temperature rises (see their figure 9).  That this 
happens to at least the $\sim$1300~\AA\ feature is consistent with the 
probable identification of much of the blend with modest-to-low ionization 
Si~{\sc iii}/O~{\sc i} transitions.  The pattern in figure~\ref{fig:comp} is 
that IX~Vel presents with the highest equivalent width in these dips, SS~Cyg 
is intermediate, whilst they are completely absent from QU~Car.
This trend operates in the same sense as the rise in relative prominence of 
the O\V\ and He\II\ lines.

Two final points of interest with regard to figure~\ref{fig:comp} are
the larger number of interstellar lines in the spectrum of QU~Car
compared to either IX~Vel or SS~Cyg, and the appearance of both the
C\IIIl1176 and C\IIl1335 lines in emission in QU~Car (albeit with a
superposed absorption component in the former).  The relatively rich
interstellar line spectrum has to do with the greater distance to
QU~Car. This is estimated to be at least $500\,$pc (Gilliland
\& Phillips 1982), compared to $96\,$pc and $166\,$pc, from 
astrometric parallaxes, for IX Vel ({\it HIPPARCOS}) and SS~Cyg
(Harrison et al. 2001), respectively.  The seeming predeliction for
carbon line emission may be due to a carbon over-abundance in the
emission line gas.  Both the question of the distance to QU~Car and
the abundances issue are focused on in a separate study (Drew et
al. 2002).

Although wind signatures are scarcely apparent in the QU~Car spectrum shown
in figure~\ref{fig:comp}, the comparison with archive \iue\ data 
(section~\ref{s:mag}) showed that they have been more apparent at other 
epochs.  However there is a clear point of contrast with objects like SS~Cyg 
and IX~Vel with regard to the maximum outflow velocities observed.  In QU~Car 
they are usually about 2000~\kms, whilst in both SS~Cyg and IX~Vel this is 
in the region of 4000~\kms.  This higher figure is indeed typical of 
HnMCV (see e.g. Prinja \& Rosen 1995).  Very crudely, 
if it is assumed these speeds scale as the rotation speeds in the accretion 
disc where presumably the wind is launched, this factor of 2 difference may 
be viewed as implying a substantially larger launch radius in QU~Car -- 
perhaps up to a factor of 4 larger.  Quantitatively this conclusion is 
somewhat dependent on both the binary inclination and accretor mass -- 
qualitatively it will be hard to avoid.  Its implication, like the impression 
of relatively high ionization, is a higher than typical mass accretion rate.

In the above we have assumed that the UV continuum emission in QU~Car
is dominated by disc accretion.  This seems uncontroversial given the
similarity between the slope of its dereddened energy distribution and
that of indisputably accretion-dominated HnMCV (see
section~\ref{s:mean}).  It is likely that all absorption features,
including those seen superposed on the C\III, C\IV\ and He\II\
emission profiles, form in the disc atmosphere.  This view is
supported by the way in which the pattern of variability seen in N\V\
and Si\IV\ absorption during the Q3 observation is also apparent in
the C\IV\ absorption component. The less variable He\II\ absorption
component can be viewed as mimicking the less variable O\V\
profile.

\begin{figure}
\vspace{9cm}
\includegraphics{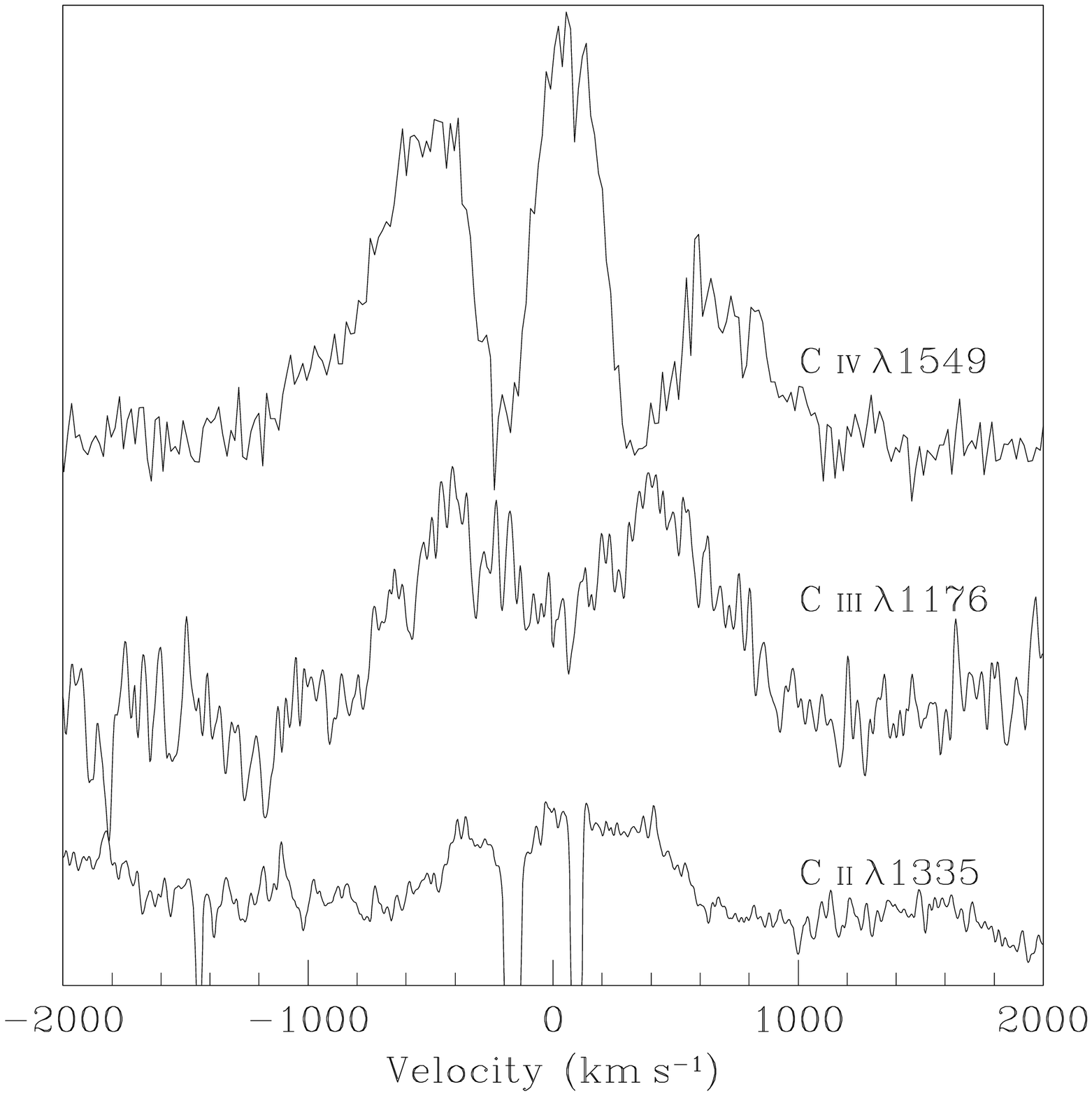}
\caption{Three ionisation stages of carbon present in the UV
spectrum: C\IIIl1176, C\IIl1335 and C\IVl1549, plotted on the same
velocity scale. The C\II\ and C\III\ lines are taken from a mean of the three STIS
spectra to improve the S/N ration, whereas the C\IV\ line is taken from I3. The narrow absorptions superposed on the C\II\ line are
interstellar lines formed in the same transition.}
\label{fig:c}
\end{figure}

We now consider the likely origin of the line emission.
Figure~\ref{fig:c} shows the emission, from the three ionization
stages of carbon present in the spectrum, plotted on a common velocity
scale.  The C\II\ line is clearly the narrowest with a HWZI of order
600\kms.  The HWZIs in the C\III\ and C\IV\ lines are not so easy to
tell apart.  On fitting a single gaussian through the line wings we
find that the FWHM, respectively for these two lines, is $\sim$950\kms
and $\sim$1100\kms.  These figures translate to HWZIs of $\sim$1200
and $\sim$1400\kms.  We do not correct for intrinsic multiplet
splitting because the magnitude of the correction depends on the
unknown flux ratios among the multiplet components.  Fortunately these
splittings are not so large that correction for them would alter the
basic finding -- namely, that broader emission is associated with
higher ionization. The near equality of the C\III\ and C\IV\ emission
widths probably indicates that the excited C\III\ line contains a
substantial contribution from recombination.

To see all three carbon ion stages simultaneously, with this pattern
of velocity widths probably points to an origin in a disc
`chromosphere'.  This might be created by irradiation from the
EUV-bright inner disc and accreting object.  Emission produced by an
outflow from the accretor or the innermost disc is rendered less
likely by the modest ionization of the C\II\ and C\III\ lines (with
the emissive flux in the C\III\ line only a little less than that in
C\IV ).  Since the emission component in each of the C\III, C\IV\ and
He\II\ lines is broader than the superposed absorption, we would have
to refer to a specific kinematic model for the system to position the
emission line region relative to the absorption line region. In the
absence of this, we cannot distinguish whether the line emission is
optically thin and overlies the source of line absorption, or whether
it comes from a different part of the disc.

The emergent picture of QU~Car is then that it is relatively highly-ionized 
compared to the more typical properties of HnMCV.  The narrowness of the
UV spectral lines observed may have to do with their production at larger
disc radii than usual.  Of course we cannot yet rule out the possibility
that QU~Car might be oriented nearly pole on. 

We have to leave as an unsolved mystery the origin of the short-term
variability within the spectral line profiles
(section~\ref{s:timevar}) that has been revealed thanks to the
HST/STIS TIME-TAG mode of data-taking.  Since this variability is not
present at all epochs and indeed presents with different properties on
the two occasions we have noticed it, it is unlikely to be connected
to anything as fundamental as the binary orbit.  Our third set of data
hints at a repetition timescale of around an hour if, that is, the
detected absorption profile disturbances do in fact repeat.  To
progress any further, another more intensive campaign of spectroscopic
observation would be required.

Lastly, we have not yet been able to begin to find an explanation for
the remarkable inverse P~Cygni character of, most notably, the
N\Vl1240 and Si\IVl1397 line profiles detected in our third dataset.
It is already clear that QU~Car is a remarkable binary that has been
overlooked for too long.  In a separate study focusing on exploitation
of the UV interstellar line data and new optical spectroscopy (Drew et
al. 2002), we show that this binary may in fact be closely related to
the supersoft sources and then present evidence of carbon
over-abundance.  It is to be hoped that this re-evaluation of the
status of QU~Car and the presentation here, of its striking UV
spectroscopic properties, will significantly broaden our perception of
the ways the Universe finds to configure such high-state compact
binaries.

\section*{Acknowledgements}
Support for Proposal number G0-8279 was provided by NASA through a
grant from the Space Telescope Science Institute, which is operated by
the Association of Universities for Research in Astronomy,
Incorporated, under NASA contract NAS5-26555

LEH would like to acknowledge the award of tuition fees and
maintenance grant provided by the Particle Physics and Astronomy
Research Council (PPARC)


\end{document}